\documentclass{PoS}

\usepackage{amsmath} 
\usepackage{graphics} 
\usepackage{braket}
\newcommand{\Slash}[1]{{\ooalign{\hfil/\hfil\crcr$#1$}}}

\title{Renormalization of two-dimensional XQCD}

\ShortTitle{Renormalization of two-dimensional XQCD}

\author{Hidenori Fukaya and \speaker{Ryo Yamamura}$^\dag$\\
        Department of Physics, Osaka University, Toyonaka, Osaka 560-0043, Japan\\
        $^\dag$E-mail: \email{ryamamura@het.phys.sci.osaka-u.ac.jp}}

\abstract{Recently, Kaplan proposed an interesting extension of QCD
named Extended QCD or XQCD with bosonic auxiliary fields~\cite{Kaplan:2013dca}.
While its partition function is kept exactly the same as that of QCD,
XQCD naturally contains properties of low-energy hadrons.
We apply this extension to the two-dimensional QCD in the large $N_c$ limit ('t Hooft model)~\cite{'tHooft:1974hx}.
In this solvable model, it is possible to directly examine the hadronic picture of the 2d XQCD
and analyze its renormalization group flow 
to understand how the auxiliary degrees of freedom behave in the low energy region.
We confirm that the additional scalar fields can become dynamical acquiring  the kinetic term,
and its parity-odd part becomes dominant in the low energy region.
This renomalization of XQCD provides an "extension" of  the renormalization scheme of QCD,
inserting different field variables from those in the original theory,
without any changes in physical observables.}

\FullConference{The 33rd International Symposium on Lattice Field Theory\\
                 14 -18 July  2015\\
                 Kobe International Conference Center, Kobe, Japan}

\begin{document}

\section{Introduction}
\label{sec:Intro}
In Ref.~\cite{Kaplan:2013dca}, Kaplan proposed an interesting reformulation of QCD named Extended QCD or XQCD.
This new formulation contains additional auxiliary bosonic fields, 
keeping the partition function of QCD unchanged.
The physics of XQCD is exactly the same as that of QCD, 
as long as the source operators of the ordinary quark and gluon fields are inserted.
Kaplan has shown that XQCD can describe several low energy hadronic pictures
such as the quark models, chiral perturbation theory and the bag models
more naturally than QCD itself.

In this work, we study the Wilsonian renormalization group (RG) flow of 
the two-dimensional version of (X)QCD (we will simply denote QCD$_2$ or XQCD$_2$ in the following),
in the large $N_c$ limit, for which we have already published a paper~\cite{Fukaya:2015cqa}.
This theory is known as the 't Hooft model~\cite{'tHooft:1974hx} and 
the advantage of studying this model is that the theory is 
particularly simplified in the large $N_c$ limit and solvable. 
We find that the auxiliary fields become dynamical
when we take into account quantum corrections.
In particular,
the (pseudo)scalar auxiliary field
should play a key role in the low energy effective action.
It contains the degrees of freedom of pions,
the lightest hadrons, as a consequence of the dynamical chiral symmetry breaking~\cite{Nambu:1961tp}.

We also find that XQCD provides an interesting extension of the renormalization ``scheme''.
Namely, we can insert an arbitrary number of new bosonic degrees of freedom at an arbitrary scale $\Lambda_{\rm cut}$  
and the RG flow is extended to the space of their new interactions which are completely absent in 
the RG flow of QCD.

\section{Extended QCD and its two-dimensional version}
\label{sec:reviewXQCD}
In this section, we review the original Extended QCD \cite{Kaplan:2013dca} 
in four dimensions and construct its two-dimensional version.
We also summarize what is known in this two-dimensional large $N_c$ QCD (the 't Hooft model).

\subsection{XQCD in four dimensions}
\label{subsec:defXQCD}
Let us consider QCD with $N_f$ flavors of quarks and gauge group $SU(N_c)$ 
in four-dimensional Euclidean spacetime.
XQCD is defined by introducing three types of auxiliary fields,
the scalar field $\Phi$, vector $\mathbf{v}_\mu$ and axial vector $\mathbf{a}_\mu$,
with the action in a Gaussian form
which keeps the original QCD partition function intact (up to a constant) :
\begin{align}
Z_{\text{QCD}}
=\int e^{-S_{\text{QCD}}[\psi,\bar{\psi},\mathbf{A}_\mu]}
=\int e^{-S_{\text{QCD}}[\psi,\bar{\psi},\mathbf{A}_\mu]
-S_{\text{aux}}[\Phi,\Phi^\dag,\mathbf{v}_\mu,\mathbf{a}_\mu]} 
\equiv
&~Z_{\text{XQCD}},
\end{align}
where the above path integration is over all fields.
Our new theory is given by
\begin{align}
\label{eq:XQCD}
S_{\text{XQCD}}
=N_c\int d^4x~\bigg[\bar{\psi}(\mathcal{D}+m)\psi 
+\frac{1}{4g^2}\text{Tr}~\mathbf{F}_{\mu\nu}\mathbf{F}_{\mu\nu}
+\lambda^2\left(\text{Tr}~\Phi^\dag\Phi
+\frac{1}{2}\text{Tr}~[{\mathbf v}_\mu{\mathbf v}_\mu
+{\mathbf a}_\mu{\mathbf a}_\mu]\right)\bigg],
\end{align}
where 
\begin{equation}
\label{eq:diracopeXQCD}
\mathcal{D}\equiv
\Slash{D}+\Slash{{\mathbf v}}+i\Slash{{\mathbf a}}\gamma_5+
2(\Phi P_++\Phi^\dag P_-).
\end{equation}
Note that $S_{\rm XQCD}$ is manifestly renormalizable.
Here, the color singlet $\Phi$ transforms as a bifundamental representation 
under the $SU(N_f)_L\times SU(N_f)_R$ chiral symmetry,
and the flavor singlets $\mathbf{v}_\mu$ and $\mathbf{a}_\mu$ 
are $N_c\times N_c$ matrices (the singlet plus
adjoint representations of the $SU(N_c)$ gauge group).
Notice that when we integrate out all auxiliary fields from the theory,
the resulting four-quark interactions are canceled by the Fierz identity:
\begin{equation}
\label{eq:fierzeu}
(P_+)_{mn}(P_-)_{m'n'}+(P_-)_{mn}(P_+)_{m'n'}=
\tfrac{1}{4}[(\gamma_\mu)_{mn'}(\gamma_\mu)_{m'n}
-(\gamma_\mu\gamma_5)_{mn'}(\gamma_\mu\gamma_5)_{m'n}].
\end{equation}
Since the integration over auxiliary fields is just a constant, 
the expectation value of any operator 
involving gluon and quark fields only, 
is equivalent to that of QCD.

\subsection{Application to the 't Hooft model}
\label{subsec:tHooft}
In this work, we consider the large $N_c$ limit of QCD in two-dimensional Lorentzian spacetime,
which is the so-called 't Hooft model~\cite{'tHooft:1974hx}.
The advantage of studying this model is that the theory is particularly simplified in the large $N_c$ limit and solvable.
In this subsection, we briefly review this model and construct its extended version.

\begin{figure}[htbp]
\begin{center}
\includegraphics[width=5cm]{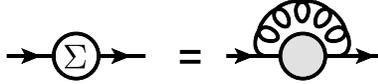}
\caption{A diagrammatic expression of the self-consistent equation for the self-energy $\Sigma$(p).}
\label{fig:selfenergy}
\end{center}
\end{figure}

It is most convenient to take the light-cone gauge: 
$\mathbf{A}_-=\mathbf{A}^+=\tfrac{1}{\sqrt{2}}(\mathbf{A}^0+\mathbf{A}^1)=0$ 
to analyze this model.
This gauge greatly simplifies the Feynman rule in the large $N_c$ limit
and enables us to non-perturbatively compute the quark self-energy.
For example, the self-energy $\Sigma(p)$ of quarks satisfies 
a self-consistent equation (see Fig.~\ref{fig:selfenergy})
whose exact solution gives the square of the constituent quark mass $M^2$ as $~M^2 = m^2-g^2/\pi$.

To ``extend'' the 't Hooft model is
almost straightforward as the original XQCD in four-dimensions with use of
the Fierz identity of two-dimensional theories.
The total action of XQCD$_2$ is given by
\begin{align}
S_{\text{XQCD}}
=
\int d^2x~\bigg[&~\bar{\psi}\mathcal{D}'\psi 
+\frac{1}{2}\text{Tr}~(\partial_-\mathbf{A}_+)^2 
\label{eq:XQCD2shift}
-\lambda^2\bigg(\text{Tr}~\Phi^\dag\Phi
-\frac{1}{2}\text{Tr}~{\mathbf v}_\mu{\mathbf v}^\mu\bigg)
\bigg],
\end{align}
where
\begin{equation}
\label{eq:diracopeXQCD2}
\mathcal{D}'\equiv
i\Slash{\partial}
-\frac{g}{\sqrt{N_c}}\mathbf{A}_+\gamma^+
+\frac{i\alpha\lambda}{\sqrt{N_c}}\Slash{{\mathbf v}}
-\frac{\sqrt{2}\alpha\lambda}{\sqrt{N_c}}
(\Phi P_++\Phi^\dag P_-),
\end{equation}
and $\lambda$ and $\alpha$ are arbitrary real parameters.
The mass dimensions of auxiliary fields and parameters are given by
$[\Phi]=[\mathbf{v}_{\mu}]=0,~[\alpha]=0,~[\lambda]=1$.

\section{Extended renormalization scheme}
\label{sec:RGQCD2}
As explained above, although QCD and XQCD are exactly equivalent,
their low energy expressions are expected to be different.
To understand this more clearly, we perform the Wilsonian 
RG transformation on both theories and 
compare their low energy effective actions.

We would like to address two possible features of XQCD.
One is how the mesonic degrees of freedom become dynamical.
As $\Phi$ is expected to play a role of the NG boson at low energy,
the RG flow should develop its kinetic term at low energy,
keeping its mass almost zero.
Another issue is to see what happens on the original quark 
and gluon sectors along the RG flow.
As hadrons play more important roles at low energy,
the original quarks and gluons should decrease their relevance,
and can eventually be decoupled from the effective action,
near the scale of their (constituent) masses.

Inclusion of the auxiliary fields extends the (relevant)
parameter space of the theory.
The new terms of the effective Lagrangian we should consider are
\begin{eqnarray}
{\rm Tr}\partial_\mu \Phi^\dagger \partial^\mu \Phi,\;\;
{\rm Tr}\partial_\nu {\bf v_\mu} \partial^\nu {\bf v}^\mu,\;\;
{\rm Tr}(\partial_\mu  {\bf v^\mu})^2, \;\;
\text{Tr}~\Phi^\dag\Phi,\;\;
\text{Tr}~{\mathbf v}_\mu{\mathbf v}^\mu,\;\;
\bar{\psi}(\Phi P_++\Phi^\dag P_-)\psi,\cdots
\end{eqnarray}
However, as the original theory has only two parameters $g$ and $m$,
the new interactions are not independent,
but essentially controlled by these two parameters.
Namely, the RG flows are restricted 
on a two-dimensional surface in the extended parameter space.

Which two-dimensional surface we take is determined by
the choice of the regularization we use, 
and the re-definition of the coupling constants 
(by giving counterterms).
Therefore, the choice of the surface corresponds to 
nothing but the choice of the renormalization scheme.
Thus, XQCD can be regarded as the extension of the renormalization
scheme to the extended theory space.
The physics remains to be unchanged
as the observables do not depend on the renormalization scheme.
The extended renormalization scheme, 
allowing the new field contents, provides us
a wider choice of the effective actions sharing the same physics.

\section{RG flow of QCD$_2$ in the large $N_c$ limit}
\label{sec:RGflow}
In this subsection, we analyze the RG flow of  QCD$_2$ itself,
without introducing any auxiliary fields.
We find a non-perturbative ``solution'' 
(in a truncated theory space), which reasonably interpolates the
theory in the continuum limit and that at the constituent quark mass.

Our goal is to integrate out the high energy modes of
the quark and gluon fields in QCD$_2$ and
obtain an effective action $S_\Lambda$ at a finite cut-off $\Lambda$.
If we could employ a gauge-invariant regularization,
we expect that $S_\Lambda$ has a similar form 
to the bare action :
\begin{align}
\label{eq:reQCD2action}
S_{\Lambda}^{\rm QCD}
&=
\int d^2x~\left[
-\frac{1}{2}\text{Tr}~(\mathbf{A}_+)_R \partial_-^2 
(\mathbf{A}_+)_R
+\bar{\psi}_R(i\Slash{\partial}-m_R(\Lambda))
\psi_R
-\frac{g_R(\Lambda)}{\sqrt{N_c}}\bar{\psi}_R\mathbf{A}_+\gamma^+\psi_R
+ \cdots\right],
\end{align}
where $(\mathbf{A}_+)_R$ and $\psi_R$ denote the renormalized fields, and
$m_R(\Lambda)$ and $g_R(\Lambda)$ are the renormalized mass and coupling constant.
In this work, we truncate the higher order terms 
and neglect irrelevant contributions at $O(1/\Lambda^4)$.

\begin{figure}[!tbhp]
\centering
\includegraphics[width=7cm]{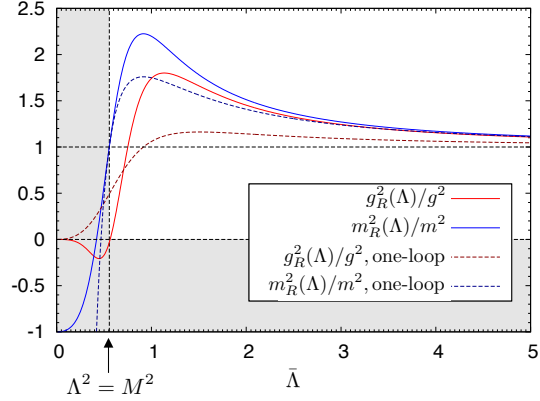}
\caption{The RG running of the mass and coupling of QCD$_2$.}
\label{fig:flow_QCD2}
\end{figure}

The renormalized mass and coupling constant are computed in Ref.~\cite{Fukaya:2015cqa}.
Here we just show the results,
\begin{align}
\label{eq:rcouplingresult}
m^2_R(\Lambda)
=m^2\left(1+
\frac{\displaystyle
\frac{2g^2}{\pi\Lambda^2}\log\bigg|\frac{\Lambda^2}{M^2}\bigg|}
{\displaystyle 
1-\frac{g^2}{\pi\Lambda^2}\log\bigg|\frac{\Lambda^2}{M^2}\bigg|}\right),~~~
g^2_R(\Lambda)
=
\frac{\displaystyle
1-\frac{g^2}{\pi\Lambda^2}}
{\displaystyle 
\left(1-\frac{g^2}{\pi\Lambda^2}\log\bigg|\frac{\Lambda^2}{M^2}\bigg|\right)^2}\hspace{0.5mm}g^2. 
\end{align}
This is the non-perturbative result at the large $N_c$ limit.
The running of them are shown in Fig.~\ref{fig:flow_QCD2}.
The solid curves are non-perturbative solutions, while the dashed ones are the one-loop results.
The flow shows that the coupling and mass do not monotonically increase but return to
near the original bare values at $\Lambda\sim M$.
Here, we make all quantities dimensionless using an arbitrarily chosen parameter $\Lambda_0$, and
use $\bar{\Lambda}=\Lambda/\Lambda_0$ for the horizontal axis.
The bare parameters are set to $g/\Lambda_0=1$ and $m/\Lambda_0=0.1$.

\section{RG flow of $\text{XQCD}_2$ in the large $N_c$ limit}
\label{sec:RGXQCD2}

Now let us investigate the RG flow of $\text{XQCD}_2$ in the large $N_c$ limit.
As in the previous section, we truncate our theory 
space to neglect $O(1/\Lambda^4)$ terms.
The large $N_c$ limit also helps to reduce some redundancy of the extended theory space.
For example, the kinetic term of $\mathbf v_\mu$ is never developed.
With this simplification, the most general form of the effective action is 
\begin{align}
S^\text{XQCD}_\Lambda= 
S^{\rm QCD}_\Lambda+\int d^2p&~\bigg[Z_{\Phi}(\Lambda)\text{Tr}~\Phi^\dag p^2\Phi
-m_\Phi^2(\Lambda)\text{Tr}~\Phi^\dag\Phi
-\frac{\sqrt{2}y(\Lambda)}{\sqrt{N_c}}
\bar{\psi}_R(\Phi P_++\Phi^\dag P_-)\psi_R \notag \\
&+\frac{1}{2}\lambda^2\text{Tr}~{\mathbf v}_\mu{\mathbf v}^\mu
+i\frac{\alpha\lambda}{\sqrt{N_c}}\frac{Z_\psi(\Lambda)}{Z_\psi(\Lambda_{\rm cut})}
\bar{\psi}_R\Slash{\mathbf{v}}\psi_R
\bigg].
\end{align}
Neglecting the overall normalization of the fields,
our theory space is extended from 2 (with $m_R$ and $g_R$)
to 5 dimensions (since $\alpha$ and $\lambda$ do not run).

As discussed in Sec.~\ref{sec:RGQCD2}, 
we can define a number of new RG schemes in this extended theory space,
by choosing a two-dimensional surface in it.
The simplest (and trivial) scheme is to take the three constraints :
\begin{align}
\label{eq:QCDscheme}
Z_\Phi (\Lambda)=0,\;\;\; m_\Phi^2(\Lambda)=\lambda^2,\;\;\; y(\Lambda)=\alpha \lambda,\;\;\;(\mbox{at any $\Lambda$}),
\end{align}
along the RG flow.
With this scheme, one can always integrate $\Phi$ and $\mathbf v_\mu$ out
and go back to original QCD$_2$ at any scale $\Lambda$.
Since this scheme is exactly equivalent to the scheme in QCD$_2$,
let us call it the ``QCD scheme''.

We are interested in more non-trivial schemes, where the hadronic degrees of freedom
become relevant (let us denote it the ``hadronization scheme'').
Let us require
the same form of the constraints as Eq.~(\ref{eq:QCDscheme})
but only at a point $\Lambda=\Lambda_{\rm cut}$:
\begin{align}
\label{eq:hadronizationscheme}
Z_\Phi (\Lambda_{\rm cut})=0,\;\;\; m_\Phi^2(\Lambda_{\rm cut})=\lambda^2,\;\;\; y(\Lambda_{\rm cut})=\alpha \lambda.
\end{align}
Then, the RG flows can go inside the bulk of the extended five-dimensional space.
In the following, we compute the RG flow of XQCD in 
this hadronization scheme and compare it with the QCD scheme.

\subsection{One-loop analysis}
\label{subsec:1loopXQCD}

\begin{figure}[tbhp]
\centering
\includegraphics[width=12cm]{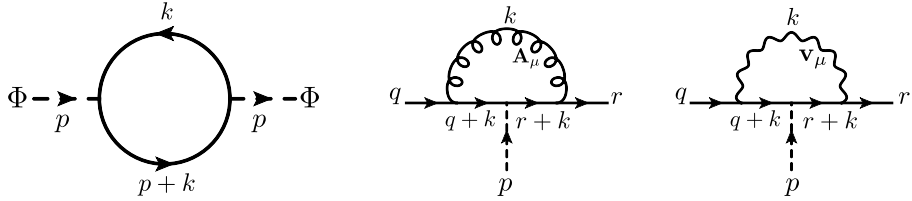}
\caption{$\Phi$'s self energy and Yukawa interaction.}
\label{fig:figs}
\end{figure}

Let us start with the computation at the one-loop.
The three relevant diagrams in the large $N_c$ limit
are the quark self-energy, 
the $\Phi$'s self energy and Yukawa interaction (Fig.~\ref{fig:figs}).

Already at this moment, we can answer to our first question
about the RG flow of the quark and gluon fields in XQCD$_2$.
The three diagrams show that the scalar (and pseudo-scalar) $\Phi$ field
receives quantum corrections from $\psi$ and ${\mathbf v_\mu}$,
but never gives a feedback to them.
Namely, the RG flow of the quark and gluon sector is unchanged.
This result is not what we originally expected :
weakening of the quark and gluon interactions.
It seems that the two-dimension, the light-cone gauge, 
and the large $N_c$ limit simplify the theory too much.
We still expect a non-trivial difference in the case of
four-dimensional QCD with $N_c=3$.

Although there is no essential change in the RG flow of the quark mass 
and gauge coupling, the Feynman diagrams are 
quite different from those in original QCD.
The essential change is in inclusion of the Yukawa interaction,
which makes the mesonic degrees of freedom more relevant,
as will be discussed below.

In Ref.~\cite{Fukaya:2015cqa}, the RG flows of the renormalized quantities at the one-loop level are computed as
\begin{align}
&Z_{\Phi}(\Lambda)
=
\frac{5y^2(\Lambda)}{6\pi}\left(\frac{1}{\Lambda^2}-\frac{1}{\Lambda_{\text{cut}}^2}\right)
+O(\Lambda^{-4}),~~
m^2_\Phi(\Lambda)
=
\lambda^2-\frac{y^2(\Lambda)}{\pi}\log\left(\frac{\Lambda_{\text{cut}}}{\Lambda}\right)
+O(\Lambda^{-2}), \notag \\
\label{eq:RGsoly}
&y(\Lambda)
=
\frac{\alpha\lambda}
{1+\frac{\alpha_R^2(\Lambda)}{\pi}\log\left(\frac{\Lambda_{\text{cut}}}{\Lambda}\right)}
+O(\Lambda^{-2}),
\end{align}
where we have used the initial conditions Eq.~\eqref{eq:hadronizationscheme}.
As expected, the $\Phi$ field becomes a dynamical
variable, developing its kinetic term.

\subsection{Non-perturbative analysis}
Here we assume that the chiral symmetry breaking in this model.
Namely the VEV of $\Phi$, which is related to the VEV of $\bar{\psi}\psi$,
takes a non-zero value. 
Thus we may re-parametrize $\Phi$ as
\begin{align}
\Phi 
=
\braket{\Phi}
e^{\frac{\sigma+i\pi}{\sqrt{2}}},~~
\braket{\Phi}=
-\frac{1}{\sqrt{2N_c}}\frac{\alpha}{\lambda}\braket{\bar{\psi}{\psi}}+O(m).
\end{align}
where $\sigma$ and $\pi$ are $N_f\times N_f$ hermitian matrices.
With this parametrization, 
the masses of $\sigma$ and $\pi$ are given by
\begin{align}
m^2_\sigma
=
\frac{\alpha^2}{N_c}\braket{\bar{\psi}\psi}^2+O(m)~,~~
m^2_{\pi}
=
\tfrac{1}{2}m\braket{\bar{\psi}\psi}+O(m^2).
\end{align}
Since the mass of $\pi$ is proportional to the quark mass, 
it vanishes in the chiral limit $m \rightarrow0$.
This GMOR relation~\cite{GellMann:1968rz} is kept along the renormalization flow as long as 
our renormalization scheme preserves the chiral symmetry.
For $\sigma$, their mass is proportional to $\Lambda^2$ 
since the mass $Z_\Phi^{-1}(\Lambda)m^2_\Phi(\Lambda)$ is proportional to $\Lambda^2$.
For the quarks, its mass is proportional to $\Lambda$ 
since the Yukawa coupling $Z_\Phi^{-1/2}(\Lambda)y(\Lambda)$ is proportional to $\Lambda$.
As we continue to integrate out  high momentum modes,
$\sigma$ and quarks would decouple from the low energy dynamics at some scale,
while $\pi$ continues to contribute to the low energy dynamics.
Eventually the theory is expected to go to the chiral effective theory 
described by the $\pi$ field only and this confirms the low energy hadronic picture. 
We never reach this picture from the RG flow without auxiliary fields.
In this way, the extension of the RG scheme introducing auxiliary fields
gives a different aspect of the theory.

\section{Summary}
\label{sec:conclusion}
We have studied the RG flow of QCD$_2$ in the large $N_c$ limit and its extension to XQCD$_2$.

For the RG flow of XQCD,
the parameter space of the theory is extended and 
the choice of the surface 
corresponds to that of the (extended) renormalization scheme. 
In ``hadronization scheme'', where the scalar auxiliary field $\Phi$ becomes dynamical 
while the vector auxiliary field $\mathbf{v}_\mu$ 
still remains to be an auxiliary,
we confirm the hadronic picture of QCD.
Due to the chiral symmetry breaking in QCD$_2$ at the large $N_c$ limit,
the only pions remain almost massless and relevant in the low energy region.
Since the RG flow of QCD does not show this picture,
we emphasize that we can never realize such a picture  
without taking into account the RG flow with auxiliary field.\\

{\bf Acknowledgements: }
We thank Kazuhiko Kamikado, David B. Kaplan, Kengo Kikuchi, Tetsuya Onogi, 
and Masatoshi Yamada for fruitful discussions and useful comments.
We also thank the Yukawa Institute for Theoretical Physics, 
Kyoto University. Discussions during the YITP workshop YITP-T-14-03 
on ``Hadrons and Hadron Interactions in QCD'' were useful to complete this work.
This work is supported in part by the Grand-in-Aid of the Japanese Ministry of Education 
No.25800147, 26247043 (H.F.), and No. 15J01081 (R.Y.).

\bibliographystyle{JHEP}
\bibliography{bib}


\end{document}